\documentclass[11pt,twoside]{article}
\usepackage{CAGN2019}
\usepackage{graphicx}

\usepackage[T1]{fontenc} 

\usepackage{latexsym}
\usepackage{verbatim}

\usepackage{ifpdf}  
\ifpdf  
      \DeclareGraphicsExtensions{.pdf,.png,.jpg}  
\else  
      \DeclareGraphicsExtensions{.eps}  
\fi 

\begin{document}

\vskip 1.0cm
\markboth{S. Zamora et al.}{Interstellar reddening correction using He I lines}
\pagestyle{myheadings}
%
%
\vspace*{0.5cm}
\parindent 0pt{Poster}


\vspace*{0.5cm}
\title{Interstellar reddening correction using He I lines}

\author{S. Zamora $^{1, 2}$, Ángeles I. Díaz,$^{1, 2}$, Elena Terlevich$^{3}$ and Vital Fernández$^{4}$}
\affil{$^{1}$Departamento de Física Teórica, Universidad Autónoma de Madrid (UAM), Spain\\
$^2$Centro de Investigación Avanzada en Física Fundamental (CIAFF), Spain\\
$^3$Instituto Nacional de Astrofísica, Óptica y Electrónica (INAOE), Mexico\\
$^{4}$Instituto de Investigación Multidisciplinar en Ciencia y Tecnología, Chile}

\begin{abstract}
We present a method to derive the logarithmic extinction coefficient in optical wavelengths using the emission lines of HeI. Using this procedure we can avoid selection biases when studying regions with different surface brightness and we can obtain better measurements of temperature lines, for example [SIII]$\lambda$6312.

\bigskip
 \textbf{Key words: } galaxies: abundances --- galaxies: ISM --- techniques: imaging spectroscopy

\end{abstract}

\section{Introduction}
Interstellar extinction is generated by the scatter and absorption of photons by the medium between the observer and the radiation source. The correction of the observed line fluxes for this effect is an indispensable preliminary step for the physical interpretation of the data. The differential extinction between observed lines can be calculated by comparing known intensity ratios of different lines which have lower dependence of physical conditions: density and temperature. It is usually derived using the hydrogen recombination line ratios for case B and assuming given values of electron density and temperature. However, the use of strong Balmer lines can lead to selection biases when studying regions with different surface brightness. This is the case in extended nebulae single exposure IFS observations, since these lines will be saturated in moderate and long exposures. 

In this work, we present a method to derive reddening corrections based only on the lines of He I, taking into account previous recombination studies to evaluate the presence of triplet states in these atoms and their influence on our recombination lines measurements \citep[][]{1972MNRAS.157..211B, 1986ApJ...310L..67F,1987MNRAS.229P..31C,1989MNRAS.238...57A,1996MNRAS.278..683S, Benjamin1999}. We have applied this procedure to calculate the reddening of different regions of 30 Dor and we quantify the differences between this method and the traditional.

\section{Data and analysis} \label{data}

\begin{figure}
\begin{center}
\hspace{0.25cm}
\includegraphics[width=0.6\columnwidth]{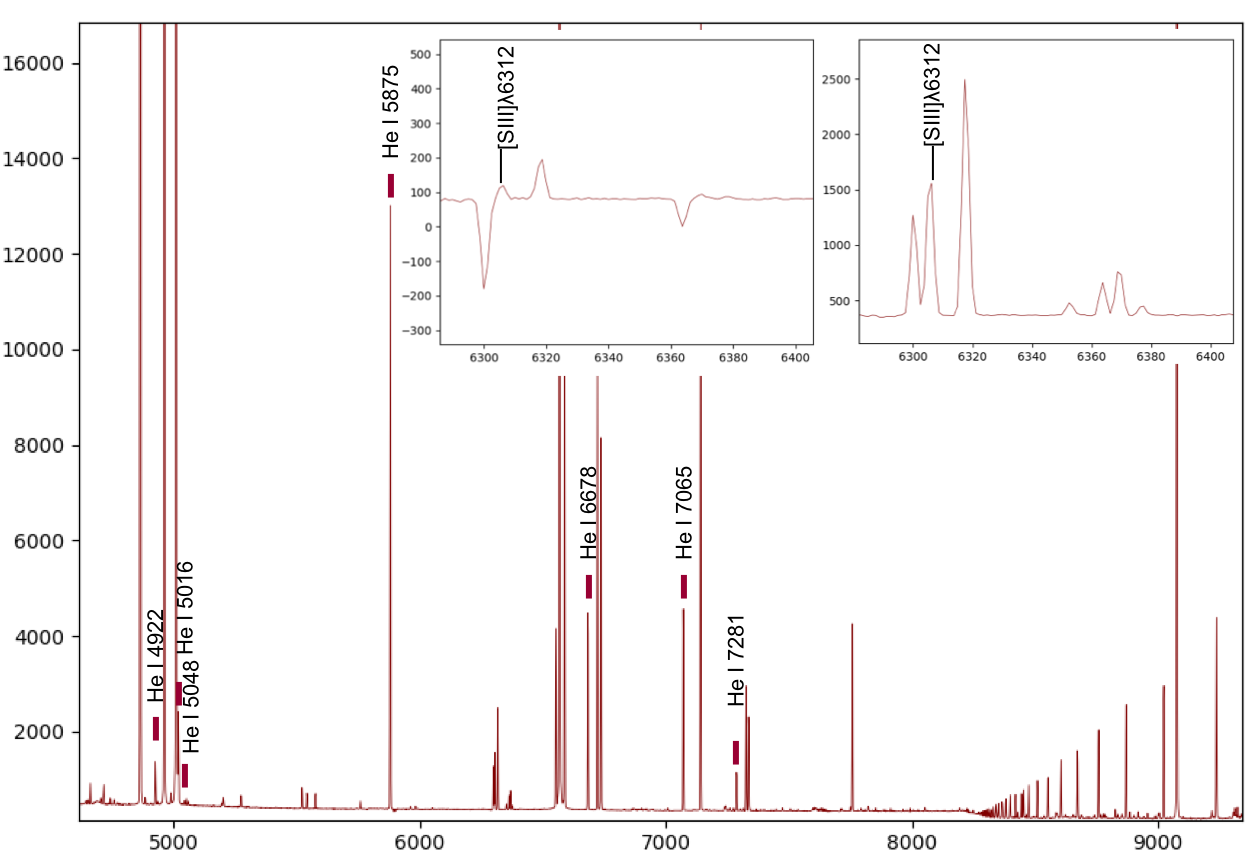}
\caption{HeI emission lines in the optical spectrum. The insets show the T$_e$ sensitive line, [SIII]$\lambda$6312, from data in the 240s (upper left panel) and 2400s (upper right panel) exposures.}
\label{Zamora2-fig1}
\end{center}
\end{figure}

We have used observations of the central part of 30Dor from MUSE \citep[Multi-Unit Spectroscopic Explorer; ][]{MUSE}. We have analysed 4 fields with exposures of 240s and 2400s from the Science Verification Program (Program ID: 60.A-93). We have selected random spectra with different surface brightness and we have measured strong lines from the cube of 240s and weak lines from the long exposure time cube. Fig.\ref{Zamora2-fig1} shows a typical spectrum of our data. These data cover a range of wavelengths from 4800 to 9300\r{A}, thus we could use all the lines shown. However, we have decided to discard: HeI$\lambda$5016, because it overlaps with [OIII]$\lambda $5007; HeI$\lambda$5048 because it is very weak and HeI$\lambda$7065, because it has very high collisional and optical depth correction  factors.

The measurement of the line fluxes has been performed using a two-component Gaussian fitting plus a linear term, to take into account simultaneously the absorption and emission in each line \citep[see ][]{Diaz2007}. Errors in the measured fluxes have been calculated from the expression given in \cite{Gonzalez-Delgado1994}. We have assumed the \cite{1972ApJ...172..593M} law with a specific attenuation of R$_v$ = 3.2, case B of recombination and a simple screen distribution of dust. The ratio of helium and hydrogen lines have been selected from \citet{pyneb}, with n$_e$(cm$^{-3}$) = 10$^2$ and T(K) = 10$^4$.

The observational flux usually is normalized using H$\beta$ as a reference line and the value of the extinction curve in this wavelength is set to null, f(H$\beta$) = 0. For the helium reddening derivation we propose to use the HeI$\lambda$6678 line for normalization. In this case:
\begin{equation}\label{eq:zamora2-2}
log_{10}\left(\frac{F_\lambda}{F_{HeI\lambda6678}}\right)-log_{10}\left(\frac{I_\lambda}{I_{HeI\lambda6678}}\right)
= -c(H\beta)(f(\lambda)-f(HeI\lambda6678))
\end{equation}
where F$_\lambda $ is the observed flux, I$_\lambda $ is the incident flux and f($\lambda$) is the extinction curve. The rest of observational and theoretical decrements ratios correspond with the dependent variable, the extinction curve represents the independent variable and we can calculate the logarithmic extinction coefficient with the slope of the fitting.

\section{Results and discussion}
\label{results}

\begin{figure}
\begin{center}
\hspace{0.25cm}
\includegraphics[width=0.65\columnwidth]{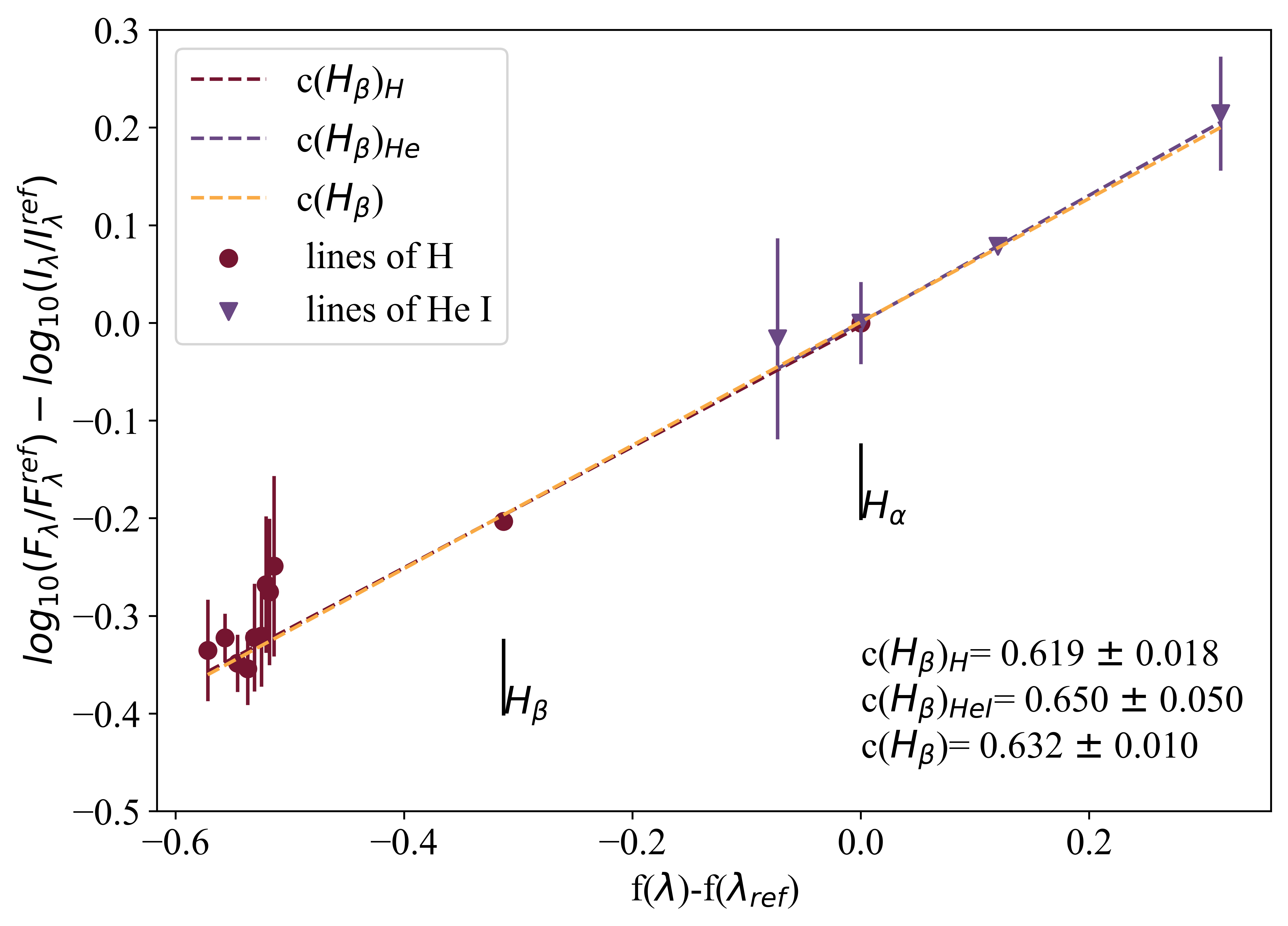}
\caption{Linear regression of c(H$\beta$) values from hydrogen lines, helium lines and both sets of lines together.}
\label{Zamora2-fig2}
\end{center}
\end{figure}

First, we have tested the spatial distribution of reference lines in our study, H$\beta$ and HeI$\lambda$6678. They have the same spatial distribution across the nebula and, also, the S/N and errors in each line are very similar in both maps. From the line measurements, we can estimate the mean associated errors. The relative error with respect to the line flux is 0.92\% for H$\beta$ and 0.15\% for H$\alpha$ in exposures of 240s. In exposures of 2400s, this value is 0.37\% for HeI$\lambda$5875 and 1.37\% for HeI$\lambda$6678. With these pairs of lines, we can calculate c(H$\beta$) for hydrogen and helium with final errors of 0.02 and 0.05 respectively. In order to minimize errors in the reddening determination, we have used the linear regression proposed in Eq. \ref{eq:zamora2-2}. Fig. \ref{Zamora2-fig2} shows an example of this method. We can see the result of the logarithmic extinction coefficient for hydrogen lines, helium lines and all lines together. Results are totally compatible among them and the intercept of the fitting is compatible with zero. Also, the error in the hydrogen determination is smaller than in that of helium. The best result is obtained using all lines simultaneously.

In the triplet levels, radiative recombinations fall to the 2$^3$S metastable level, thus the collisional contribution of lines is an important aspect. The critical density of this level is 3.3$\times $10$^3$ cm$^{-3}$ at 20000K \citep{Osterbrock2006}, less than the expected density for a typical HII regions. However, at n < n$_c$(2$^3$S), collisional contributions are significant to the 7067 \AA\ line \citep[15.2\% at 20000K;][]{Benjamin1999} and the effect of optical depth on this level is also important for this line\citep[][]{Benjamin2002}. Nevertheless, the collisional contribution to the lines proposed in this study, after discarding HeI$\lambda$7065, is within the measurement errors and their effect is minimized in the case of a linear fit. The same happens with the effects of radiative transfer and optical depth on some levels. In the measurement of helium lines, the underlying stellar absorption \citep[see ][]{2021JCAP...03..027A} is not increased with respect to the case of hydrogen. If this aspect were a critical issue in our work, we can perform the fitting to the lines following the method proposed by \cite{1994ApJ...435..647I}. Additionally, the ratios of HeI lines are less affected by temperature and density variations \citep{pyneb} than for the H$\alpha$/H$\beta$ ratio. 
 
The comparison between helium and hydrogen determinations is totally compatible and in long exposures, such as observations with saturated H$\alpha$ lines, we can use H$\beta$, Paschen lines, and Helium lines mentioned in this study to calculate reddening corrections. In conclusion, with this methodology, we can avoid selection biases when studying regions with different surface brightness in, for example, IFS observations, and we can do a complete study using only weak emission lines.

\acknowledgments This research has made use of the services of the ESO Science Archive Facility  
and it has been supported by Spanish grants AYA2016-79724-C4-1-P, PID2019-107408GB-C42 and BES-2017-080509. 

\bibliographystyle{aaabib}
\bibliography{Zamora2}

\begin{thebibliography}{}

\bibitem[\protect\astroncite{{Almog} \& {Netzer}}{1989}]{1989MNRAS.238...57A}
{Almog} Y., {Netzer} H., 1989,
\newblock {\em \mnras}, {\bf 238}, 57

\bibitem[\protect\astroncite{{Aver} et~al.}{2021}]{2021JCAP...03..027A}
{Aver} E., {Berg} D.~A., {Olive} K.~A., {Pogge} R.~W., {Salzer} J.~J.,
  {Skillman} E.~D., 2021,
\newblock {\em \jcap}, {\bf 2021(3)}, 027

\bibitem[\protect\astroncite{{Bacon} et~al.}{2010}]{MUSE}
{Bacon} R., {Accardo} M., {Adjali} e.~a., 2010,
\newblock {\em {The MUSE second-generation VLT instrument}}, Vol. 7735 of {\em
  Society of Photo-Optical Instrumentation Engineers (SPIE) Conference Series},
  p. 773508

\bibitem[\protect\astroncite{{Benjamin} et~al.}{1999}]{Benjamin1999}
{Benjamin} R.~A., {Skillman} E.~D., {Smits} D.~P., 1999,
\newblock {\em \apj}, {\bf 514(1)}, 307

\bibitem[\protect\astroncite{{Benjamin} et~al.}{2002}]{Benjamin2002}
{Benjamin} R.~A., {Skillman} E.~D., {Smits} D.~P., 2002,
\newblock {\em \apj}, {\bf 569(1)}, 288

\bibitem[\protect\astroncite{{Brocklehurst}}{1972}]{1972MNRAS.157..211B}
{Brocklehurst} M., 1972,
\newblock {\em \mnras}, {\bf 157}, 211

\bibitem[\protect\astroncite{{Clegg}}{1987}]{1987MNRAS.229P..31C}
{Clegg} R.~E.~S., 1987,
\newblock {\em \mnras}, {\bf 229}, 31P

\bibitem[\protect\astroncite{{D{\'\i}az} et~al.}{2007}]{Diaz2007}
{D{\'\i}az} {\'A}.~I., {Terlevich} E., {Castellanos} M., {H{\"a}gele} G.~F.,
  2007,
\newblock {\em \mnras}, {\bf 382(1)}, 251

\bibitem[\protect\astroncite{{Ferland}}{1986}]{1986ApJ...310L..67F}
{Ferland} G.~J., 1986,
\newblock {\em \apjl}, {\bf 310}, L67

\bibitem[\protect\astroncite{{Gonzalez-Delgado}
  et~al.}{1994}]{Gonzalez-Delgado1994}
{Gonzalez-Delgado} R.~M., {Perez} E., {Tenorio-Tagle} G., {Vilchez} J.~M.,
  {Terlevich} E., {Terlevich} R., {Telles} E., {Rodriguez-Espinosa} J.~M.,
  {Mas-Hesse} M., {Garcia-Vargas} M.~L., {Diaz} A.~I., {Cepa} J., {Castaneda}
  H., 1994,
\newblock {\em \apj}, {\bf 437}, 239

\bibitem[\protect\astroncite{{Izotov} et~al.}{1994}]{1994ApJ...435..647I}
{Izotov} Y.~I., {Thuan} T.~X., {Lipovetsky} V.~A., 1994,
\newblock {\em \apj}, {\bf 435}, 647

\bibitem[\protect\astroncite{{Luridiana} et~al.}{2015}]{pyneb}
{Luridiana} V., {Morisset} C., {Shaw} R.~A., 2015,
\newblock {\em \aap}, {\bf 573}, A42

\bibitem[\protect\astroncite{{Miller} \& {Mathews}}{1972}]{1972ApJ...172..593M}
{Miller} J.~S., {Mathews} W.~G., 1972,
\newblock {\em \apj}, {\bf 172}, 593

\bibitem[\protect\astroncite{{Osterbrock} \& {Ferland}}{2006}]{Osterbrock2006}
{Osterbrock} D.~E., {Ferland} G.~J., 2006,
\newblock {\em {Astrophysics of gaseous nebulae and active galactic nuclei}}

\bibitem[\protect\astroncite{{Smits}}{1996}]{1996MNRAS.278..683S}
{Smits} D.~P., 1996,
\newblock {\em \mnras}, {\bf 278(3)}, 683

\end{thebibliography}

\end{document}